\documentclass[pre,aps,groupaddress,showpacs,twocolumn]{revtex4-1}
\usepackage{epsfig,amsmath,graphicx,amssymb,overpic}
\usepackage{colordvi}
\usepackage{color}
\def\be{\begin{equation}}
\def\ee{\end{equation}}
\def\bee{\begin{eqnarray}}
\def\ene{\end{eqnarray}}
\def\bes{\begin{subequations}}
\def\ees{\end{subequations}}

\newcommand{\cE}{{\cal E}}
\newcommand{\br}{{\bf r}}

\newcommand{\pD}{\partial D}

\newcommand{\hvarphi}{ \hat{\varphi}}
\newcommand{\heta}{ \hat{\eta}}

\begin{document}
\title{Two-Dimensional Superfluid Flows in Inhomogeneous Bose-Einstein Condensates}

\author{Zhenya Yan$^{1}$}
\email{zyyan\_math@yahoo.com}

\author{V. V. Konotop$^{2}$}

\author{A. V. Yulin$^{2}$}

\author{W. M. Liu$^{3}$}

\affiliation{$^1$Key Laboratory of Mathematics Mechanization, Institute of
Systems Science, AMSS, Chinese Academy of Sciences, Beijing
100190, China
\\
$^2$Centro de F\'isica Te\'orica e Computacional and Departamento de F\'isica, Faculdade de Ci\^encias,
Universidade de Lisboa,  Avenida
Professor Gama Pinto 2, Lisboa 1649-003, Portugal\\
$^{3}\!$Beijing National Laboratory for Condensed Matter Physics, Institute of Physics,
 Chinese Academy of Sciences, Beijing
100190, China}


\begin{abstract}

\vspace{0.05in}
We report a novel algorithm of constructing linear and nonlinear potentials in
the two-dimensional Gross-Pitaevskii   equation subject to
given boundary conditions, which allow  for exact analytic
solutions. The obtained solutions represent
superfluid flows in inhomogeneous  Bose-Einstein condensates. The method is based on the combination of the
similarity reduction of the two-dimensional Gross-Pitaevskii equation to
the one-dimensional nonlinear Schr\"odinger equation, the latter allowing for
exact solutions, with the conformal mapping of the given domain, where
the flow is considered, to a half-space. The stability of the
obtained flows is addressed. A number of stable and
physically relevant examples are described.

\end{abstract}


\maketitle


\section{Introduction}

Nowadays one observes  rapidly increasing interest in studying   nonlinear Schr\"odinger (NLS) equations with inhomogeneous  coefficients and, in particular, in obtaining their exact analytical solutions for the physically relevant statements. Starting with the first results on integrable inhomogeneous models~\cite{integrable}, this activity received further development due to its relevance in nonlinear optics~\cite{Serkin} and in the mean-field theory of Bose-Einstein condensates (BECs)~\cite{BEC,TGK}. A possibility of constructing exact solutions was also reported for  NLS equations with inhomogeneous complex-valued coefficients~\cite{Abdul}.
However, there are two main limitations of the presently available results. First, most of them were obtained for
one-dimensional or quasi-one-dimensional statements and only a few results on  multidimensional  problems were reported, so far~\cite{TGK,YK}. Second, the most of the models allowing for construction of exact solutions were posed in the infinite domains. The first suggestion of an algorithm for constructing exact solutions of the one-dimensional (1D) NLS equation on a half-line, modeling a BEC interacting with a rigid surface, was recently reported in~\cite{BYK}.

Here we show that the  requirements of one-dimensionality and unboundness of the domain can be removed and exact analytical solutions can be obtained for models defined on bounded 2D domains and described by the NLS equation with inhomogeneous linear, $V_{\rm ext}({\bf r})$, and nonlinear, $g({\bf r})$,  potentials (the both being real-valued functions of the spatial coordinates). Moreover, some of the reported solutions are found to be stable, and thus having particular physical relevance.

The paper is organized as follows. In Sec. II, we present the 2D physical model and give the conformal mapping to reduce the 2D physical model subject to
given boundary conditions to nonlinear ordinary differential equation solved. In Sec. III, we concentrate on the two simplest representative examples to illustrate our novel method. Sec. IV is devoted to the numerical simulations for the solutions obtained in the domain $D_2$. In Sec. V, we study the generalization of the conformal mapping and present exact solutions of the 2D physical model. Finally, the outcomes are summarized in the
 Conclusion.

\section{The physical model and conformal mapping}

To be specific, we deal with the 2D nonlinear physical model
\begin{eqnarray}
\label{nls}
 i \partial_t\!\Psi(\br,t)\! =\! \left[-\frac{1}{2}\nabla^2 \!+\! V_{\rm ext}(\br)
 \!+\! g(\br)|\Psi(\br,t)|^{2}\right]\!\!\Psi(\br,t),\,\,
 \end{eqnarray}
where $\br\equiv(x,y)\in D\subset\mathbb{R}^2$, $D$ is an open domain,
and $\nabla   \equiv\left(\partial_x, \partial_y\right)$.
We are particularly interested in applications of our results to BEC flows, where $\Psi(\br, t)$ is the macroscopic wavefunction and  model (\ref{nls}) is also termed, the Gross-Pitaevskii (GP) equation~\cite{GP}. We explore the flexibility of potentials in the BEC applications, i.e., possibility of manipulating them by external electric and/or magnetic fields for the sake of creation of desirable spatial configurations for the linear potential  and for the scattering length of the two-body interactions (the latter performed through the Feshbach resonance technique~\cite{Feshbach}). We also notice that the model (\ref{nls}) has also direct relevance to the mean-field theory of exciton-polariton condensates~\cite{Amo}. There on the one hand, the 2D statement, i.e. the statement considered in this paper, is the most typical one. On the other hand applying the external pump from the free edges of a specimen one can create different kinds of nonzero conditions (nonzero currents, as required bellow in the present paper).

We concentrate on a BEC in a domain $D$ bounded by impenetrable walls. Respectively Eq.~(\ref{nls}) will be supplied by the
zero  conditions given at the boundary of the domain $D$,
which we denote as $\pD$, i.e., we impose $\Psi(\br)=0$ for all  $\br\in \pD$.

 Our goal is to find an algorithm allowing for
systematic constructions of the linear, $V_{\rm ext}(\br)$,  and nonlinear, $g(\br)$, potentials,
 for which the formulated  Dirichlet problem  allows for  exact analytical solutions.
To this aim we assume that there exists a complex analytic function
\bee
\label{cm}
\zeta(\br)\equiv \eta(\br)+i\varphi(\br)=f(z)
\ene of the complex variable
$z=x+iy\in\mathbb{C}$,
which provides the conformal mapping of the contour $\pD$  to the imaginary axis of $\zeta(\br)$, i.e., to $\eta(\br)=0$,
such that the domain $D$ is mapped into the right half-plane in terms of the new variables $(\eta,\varphi)$: $\eta(\br)>0$.
Due to analyticity of the   mapping $\zeta(\br)$ the Cauchy-Riemann equations on the pair of real-valued functions
$\eta(\br)$ and $\varphi(\br)$ hold:
\bee \label{CR}
 \partial_x\eta(\br)=\partial_y\varphi(\br),\quad \partial_y\eta(\br)=-\partial_x\varphi(\br).
 \ene
Respectively, the following constraints
 \bee
 \nabla^2\eta(\br)=0,\,\,\,\, \nabla^2\varphi(\br)=0,\,\,\,\, \nabla\eta(\br)\!\cdot\!\nabla\varphi(\br)=0
 \label{condition} \ene
 are verified, as well. In other words, $\eta(\br)$ and $\varphi(\br)$ are both 2D
  harmonic functions with orthogonal gradients.
  Moreover, on the basis of the Cauchy-Riemann  equations we find the relation
  $|\nabla\eta(\br)|^2\equiv|\nabla\varphi(\br)|^2$.

We restrict the consideration to the linear and nonlinear potentials which can be  represented in terms of  $\eta(\br)$  as follows
 \bee \label{potential}
 V_{\rm ext}(\br)\equiv -\frac{\varepsilon}{2}\,|\nabla\eta(\br)|^2,
 \quad
 g(\br)\equiv \frac{\mathcal{G}}{2}\,|\nabla\eta(\br)|^2
\ene
where $\varepsilon$ and $\mathcal{G}$ are real parameters. Without loss of generality, we choose $\mathcal{G}=\pm1$. The parameter $\varepsilon$ determines the proportionality coefficient between the potentials: $V_{\rm ext}(\br)/g(\br)=-\varepsilon/\mathcal{G} $. Then
the change of the dependent variables $\Psi(\br,t)\to \psi(\eta,\varphi,t)$,
allows one to reduce Eq.~(\ref{nls}) to the 2D form
\bee \label{stab_NLS2}
i\partial_t\psi=\varrho(\eta,\varphi)
\left(-\partial_{\eta}^2-\partial_{\varphi}^2-\varepsilon+\mathcal{G}|\psi|^2\right)\psi.
 \ene
Here  $\varrho(\eta,\varphi)$ is the positive definite
function, defined as $\varrho(\eta,\varphi)=|\nabla\eta|^2/2\equiv |\nabla\varphi|^2/2$, where the gradients must be expressed
in terms of $\eta$ and $\varphi$ (see the examples below).   The obtained equation (\ref{stab_NLS2}) is considered for $\eta>0$ and has to be supplied
with the zero boundary condition $\psi(\eta=0,\varphi,t)=0$.

First, we concentrate on time-independent solutions of Eq.~(\ref{nls}): $\Psi(\br,t) \equiv
\psi(\eta,\varphi)$, where   $\psi(\eta,\varphi)$ solves the
2D stationary GP equation with constant coefficients
$\varepsilon\psi=-(\partial_{\eta}^2+\partial_{\varphi}^2)\psi+\mathcal{G}|\psi |^2\psi$.
Particular solutions of this equation can be represented  as
\begin{eqnarray} \label{ansatz}
\psi(\eta,\varphi) = e^{i\nu \varphi}\phi(\eta ),
\end{eqnarray}
where
$\nu$ is a constant and the real-valued function $\phi(\eta)$ solves the
 problem
\bee {\cal
E}\phi(\eta)=-\partial_{\eta}^2\phi(\eta)+\mathcal{G}|\phi(\eta)|^2\phi(\eta), \quad
\phi(0)=0,
\label{ODE}
\ene
with ${\cal E}=\varepsilon-\nu^2$ and $\eta>0$.

Turning to  the physical meaning of the obtained solutions, we observe that it follows from the ansatz, $\psi(\eta,\varphi) = e^{i\nu \varphi}\phi(\eta )$,  that $ \nabla\varphi(\br)$ can be identified as the superfluid velocity. Hence, the introduced analytic function $f(z)$
is nothing but the complex potential of the respective two-dimensional flow.
 As it is well known~\cite{hydro}   such a potential defines the current, $J_C$ of the fluid through a given contour
  $C\subset D$, as well as the circulation $\Gamma_C$ along $C$:
\bee
\int_Cf^\prime(z)dz=\Gamma_C +iJ_C
\ene
(the prime stands for the derivative with respect to  $z$).
Since in our case $f(z)$ is analytic, this integral is zero for any closed contour $C$   bounding a simply connected domain. Thus the described flow has neither sources nor vorticity in $D$.

We also observe that if the change of variables implies growth of $|\nabla\varphi(\br)|$ with $\br$, then the physical meaning might have only solutions with densities decaying at the infinity (thus ensuring decaying currents). Whenever one concerns with finite densities at the infinity, the physically meaningful solutions would correspond to  $\nu=0$. This last constraint is assumed in what follows.

Finally, we notice that if the contour  $\partial D$ included in the proposed scheme is closed, this implies that  linear and (or) nonlinear potentials are divergent at some point(s) of the boundary. This is clear from the nature of the conformal mapping, since in this case there should be a point of the boundary, which is mapped into the infinity point. In its turn such a point gives an origin to the singularity of the linear and nonlinear potentials. Such cases will be exclude in what follows, although they still may have physical relevance.

\section{Examples of exact solutions}

While large diversity of the domains can be considered, here we concentrate on the two simplest representative examples for the conformal mapping (\ref{cm}).

  \subsection{The quadrant $x>|y|$ with the boundary $x=\pm y$}

The first domain  considered below, is  given by $D_1=\{ x>|y|\}$ (i.e., $D_1$ is the  quadrant of the $(x,y)-$plane) with the boundary $\partial D_1=\{x=y,\,\,y>0\}\cup \{x=-y, y<0\}$. Respectively, the
 conformal mapping is chosen as $f(z)=z^2$, and thus $\eta(\br)=x^2-y^2$ and $\varphi(\br)=2xy$.
The linear and nonlinear potentials are now given by
\begin{eqnarray}\label{pot1}
V_{\rm ext}(\br)=-2\varepsilon|\br|^2,\qquad g(\br)=2\mathcal{G}|\br|^2,
\end{eqnarray}
i.e., they are the linear  expulsive parabolic potential and the parabolic nonlinearity. Now $\varrho(\eta,\varphi)=2\sqrt{\eta^2+\varphi^2}$.

\subsection{The strip $0<y<\pi$ with the boundary $y=0,\pi$ }

Another domain explored below is a strip $D_2=\{x\in\mathbb{R},\, 0<y<\pi\}$ with the boundary
$\partial D_2=\{x\in\mathbb{R},\,y=0\}\cup \{x\in\mathbb{R},\, y=\pi\}$. Now  the function performing
 conformal mapping to the upper half-plane is $f(z)=e^z$ and the new variables are determined as $\eta(\br)=e^x\sin y$
  and $\varphi(\br)=e^x\cos y$. The linear and nonlinear potentials allowing for the exact solutions are now given by
\begin{eqnarray} \label{pot2}
V_{\rm ext}(\br)=-2\varepsilon e^{2x},\qquad g(\br)=2\mathcal{G}e^{2x},
\end{eqnarray}
and $\varrho(\eta,\varphi)=\frac 12 \left(\eta^2+\varphi^2\right) $.

\subsection{Exact solutions}

\begin{figure}
\begin{center}
{\scalebox{0.4}[0.42]{\includegraphics{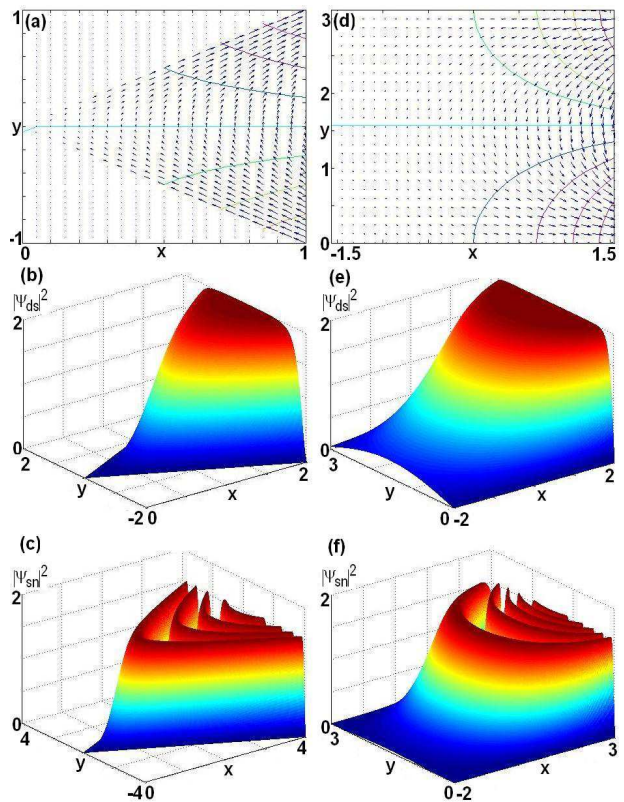}}}
\end{center}
\vspace{-0.15in} \caption{(color online). The profiles for exact solutions given by Eqs.~(\ref{ds}) and (\ref{sn}) and the corresponding velocity fields in the domains $D_1$ and $D_2$.
  Left column: Panels (a), (b), (c) show the velocity field $\nabla\varphi(\br)=(2y, 2x)$, and densities $|\Psi_{ds}(\br)|^2$ and $|\Psi_{sn}(\br)|^2$ for solutions  given by (\ref{ds}) and (\ref{sn}) in the domain $D_1$. Right column: Panels (d), (e), (f) show the velocity field $\nabla\varphi(\br)=(e^x\cos y, e^x\sin y)$, and densities $|\Psi_{ds}(\br)|^2$ and $|\Psi_{sn}(\br)|^2$  for solutions   given by~(\ref{ds}) and (\ref{sn}) in the domain $D_2$. The parameters are  $\varepsilon=2$, $\nu=0$, and $k=0.8$. }
\label{fig1}
\end{figure}

Turning to the exact solutions, below we consider only the case of repulsive interactions  $\mathcal{G}=1$, as the most natural candidate to produce stable stationary flows.  As the two simplest solutions of Eq.~(\ref{nls}) (see e.g., Refs.~\cite{YK,FT}) we study, the ``dark soliton" shape
\bee
\label{ds}
\Psi_{ds}({\br})=\psi_{ds}(\eta,\varphi)
=\sqrt{{\cal E}}\tanh\!\left( \sqrt{\frac{{\cal E}}{2}}\,\eta (\br)\right)e^{i\nu\varphi(\br)} \quad
\ene
 and the nonlinear periodic modulation
\bee\label{sn}
\Psi_{sn}({\br})\!=\!\psi_{sn}(\eta,\varphi)
\!= \!\frac{k\sqrt{2{\cal E}}}{\sqrt{1+k^2}}\,{\rm sn}\!\!
\left(\!\!\frac{\sqrt{{\cal E}}\,\eta(\br)}{\sqrt{1+k^2 }}, k\!\right)\!e^{i\nu\varphi(\br)}\,\,\,\,\,
\ene
 with ${\cal E}=\varepsilon-\nu^2>0$ and $k\in(0,\,1]$ being the modulus of the Jacobi elliptic
 sn--function [notice that $\psi_{ds}(0,\varphi)=\psi_{sn}(0,\varphi)=0$, in conformity with the imposed boundary conditions]. These exact solutions given by Eqs.~(\ref{ds}) and (\ref{sn}) and the corresponding
 velocity fields in the domains $D_1$ and $D_2$ are illustrated in Fig.~\ref{fig1}.

The obtained exact solutions, however leave several open questions related to their practical feasibility. First,  the  stability of the flows was not investigated, so far. Second, in all considered cases  infinitely growing potentials were used, while any cut-off (which exists in the real world) may strongly perturb, and even destroy the solutions. To address these issues we now turn to direct numerical simulations.

In the case at hand the potentials $V_{\rm ext}(\br)$ and $g(\br)$   grow  with $x$, while  the density goes asymptotically to (or is bounded by) a certain constant level. Bearing this in mind we   construct a physical system with bounded potential in the following way. For $x<0$ we consider  shifted linear  and nonlinear potentials $V_{\rm ext}( \br-\br_0)$ and $g(\br-\br_0)$, where $\br_0=(x_0,0)$ with $x_0$ being a constant shift vector, while $V_{\rm ext}(\br)$ and $g(\br)$ are given by the analytical formulas (\ref{potential}). To define the physical potentials at $x>0$ we mirror them  at $x=0$, thus obtaining
$V_{\rm ext}( -\br+\br_0)$ and $g(-\br+\br_0)$.  To confine the condensate along $y$-coordinate we introduce additional linear tarp potential as follows: $v_{add}=0$ for $\br \in D$ and $v_{add}=V_0$, where $V_0$ is large enough, for $\br \notin D$. For more details see the next chapter where we discuss the numerical studies of the problem.

\section{Direct numerical simulations}

\begin{figure}
\begin{center}
\includegraphics[width=\columnwidth]{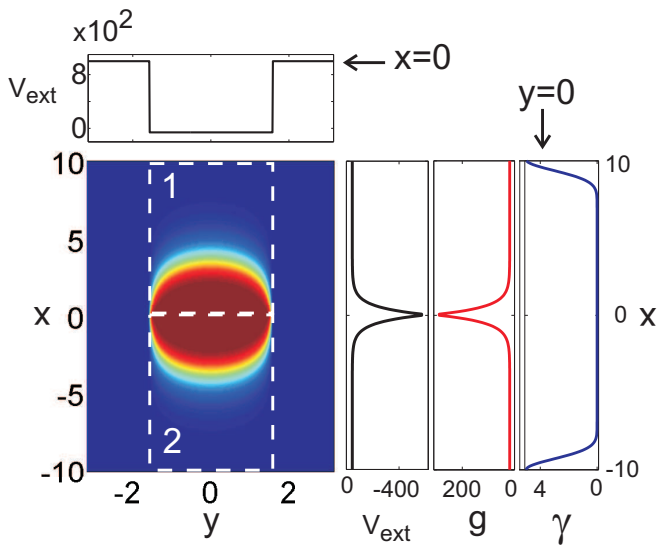}
\end{center}
\vspace{-0.15in} \caption{(color online). The ansatz produced from the solutions the with parameters  $\varepsilon=2$, $\nu=0$. In the upper part of the figure the distribution of the linear potential along $y$ is shown for $x=0$. On the right the distributions of the linear and nonlinear potentials are shown alongside with the introduced loss for $y=0$. In the areas marked by ``1" and ``2" the anzatz exactly coincides with the analytical solution (neglecting exponentially weak loss). }
\label{A_fig_new}
\end{figure}

Now we can perform  direct numerical simulations of Eq.~(\ref{nls}) with the additional confining potential $v_{add}$ taking the  initial distribution of the field in the form given by the analytical formulas $\psi(\br-\br_0)$ for $x<0$ and $\psi(\br_0-\br)$ for $x>0$. As it is clear our ansatz does not satisfy  Eq.~(\ref{nls}) only along the line $x=0$ and in the areas $y<0$ and $y>\pi$. Since, however we impose strong confining potential, we  expect that the field is very weak outside the stripe $0<y<\pi$ and our ansatz stays sufficiently close to the real stationary solution of Eq.~(\ref{nls}) with the introduced physical potentials. The analytically found solution and the corresponding potentials  are shown in Fig.~\ref{A_fig_new}.  Within the areas ``1" and ``2" the ansatzes used in the numerics as the initial condition coincides with the analytical solutions. However on the boundary between these areas the ansatzes do not satisfy the equation for stationary fields. The ansatz does not satisfy the equation outside the areas either. However the density of the condensate is low outside the areas and so there are some reasons to believe that the ansatz is close to the stationary solution. Let us remind here that in our numerical simulations to keep the condensate localized within the stripe we used strong linear potential, see the distribution of the linear potential along $y$.

\begin{figure}
\begin{center}
\includegraphics[width=\columnwidth]{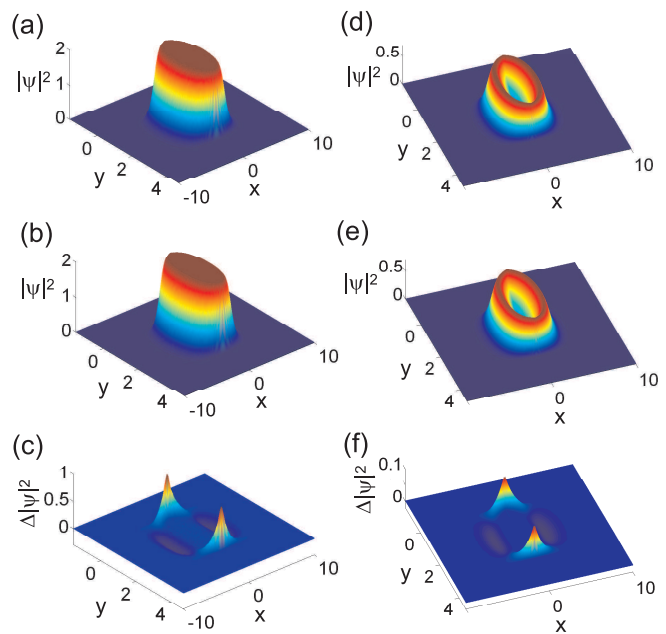}
\end{center}
\vspace{-0.15in} \caption{(color online). Panels (a) and (d) show the initial distribution of the density of the condensates for cases 1 and 2; panels (b) and (e) show the distributions of the condensate densities at $t=100$; panels (c)and (f) show the differences between the initial density distributions and the density distributions at $t=100$. The shifts of the initial conditions and the potentials are $x_0=-2.5$ for (a)-(c) and $x_0=-2$ for (d)-(f). The solution parameters are $\varepsilon=2$, $\nu=0$. The field is kept within the stripe by strong repelling linear potential $v_{a}=10^4$ for $y<0$ and $y>\pi$. The parameters for the linear losses are $\gamma_0=5$, $w_0=1$, $x_l=10$.}
\label{A_fig}
\end{figure}

We carried out numerical simulations for the solutions obtained in the domain $D_2$ and found out that some non-stationary excitations appear but the solution survive and stay rather close to the initial field distribution. To check the stability of the solution for longer time we had to get rid of the propagating excitations. Since we  perform numerical simulations  in finite windows the only way to eliminate the propagating excitations is to introduce losses in the area where initially the density of the condensate is exponentially weak. To do this we introduce linear losses in the form $\gamma=\gamma_0 \left[ \exp(-(x-x_l)^2/w_0^2)+ \exp(-(x+x_l)^2)/w_0^2\right]$. Then the losses are negligible in the area where our analytical solution is big and so the solution is practically unaffected by the artificial losses used in the numerics. In the same time all the perturbations propagating away are quickly absorbed.  In this case the excitations disappear very quickly and one can see in Fig.~\ref{A_fig} that the surviving solution is very close to the initial distribution. As it is  expected the stationary solution deviates from the ansatz most strongly around the points ($x=0$, $y=0$) and ($x=0$, $y=\pi$).

 In the  case of attractive interactions (of a negative scattering length) the solution appears to be unstable against collapse.

\section{The generalized conformal mapping}

Next  we address   possibilities of getting more general types of the potentials, allowing for exact solutions.
To this end we introduce the  generalized relations [c.f. the Cauchy-Riemann equations (\ref{CR})]:
\begin{eqnarray}\label{GCR}
\rho^2(\heta)\partial_x\heta(\br)=\partial_y\hvarphi(\br), \,\,\, \rho^2(\heta)\partial_y\heta(\br)=-\partial_x\hvarphi(\br),
\end{eqnarray}
where $\rho^2(\heta)$ is a positive-definite function of $\heta(\br)$ only [as it is clear the arguments presented below can be also applied to the case where $\rho^2(\hvarphi)$ is a function of $\hvarphi(\br)$ only].  These equations still define transformation to the orthogonal coordinates $(\heta,\hvarphi)$, however now satisfying the relations [c.f. Eq.~(\ref{condition})]
 \bee
  \nabla\!\cdot\![\rho^2(\heta)\nabla\heta]=0, \, \,
  \nabla\!\cdot\![\rho^2(\heta)\nabla\hvarphi]=0,\, \,
   \nabla\heta\!\cdot\!\nabla\hvarphi=0. \quad
\label{setg}
\ene
Moreover, now we have that
  $|\nabla\hvarphi(\br)|^2\equiv \rho^4(\heta)|\nabla\heta(\br)|^2$. Notice that the generalized case [c.f. Eq.~(\ref{setg})] can be reduced to the case mentioned above [c.f. Eq.~(\ref{condition})] in the special case $\rho^2(\heta)\equiv 1$.

Next we  repeat the steps described above for the conformal mapping, and by the direct algebra show that the generalized ansatz
\begin{equation}
\psi(\heta,\hvarphi)=\rho(\heta)e^{i\nu\hvarphi(\br)} \phi[\heta(\br)]
\end{equation}
solves the 2D stationary GP equation (\ref{nls}) provided that $\phi(\br)$  satisfy the stationary
  equation (\ref{ODE}) with $\eta(\br)$ substituted by $\heta(\br)$ and the  linear and nonlinear potentials are given by
\bee
\label{gpotential}
\begin{array}{l}
   V_{\rm ext}(\br)\equiv\displaystyle\frac{\nabla^2\rho}{2\rho}+\frac{\nu^2(1\!-\!\rho^4)
  -\varepsilon}{2}|\nabla\heta|^2,\,\, \vspace{0.1in}\cr
 g(\br)\equiv \displaystyle\frac{\mathcal{G}|\nabla\heta|^2}{2\rho^2}.
\end{array}
\ene
Notice that now the linear and nonlinear potentials are not proportional to each other any more [c.f. Eq.~(\ref{potential})].

To construct a particular example,  we choose
\begin{equation}
\label{gcon2}
\heta(\br)\!=\!2\eta(\br)\!+\!\frac{\eta^2(\br)}{2}, \,  \hvarphi(\br)\!=\!\varphi(\br),\, \rho(\heta)\!=\frac{1}{\sqrt{2\!+\!\eta(\br)}}
\end{equation}
 with $\eta(\br)$ and $\varphi(\br)$ solving Eq.~(\ref{condition}). One can  ensure that this choice also satisfies the conditions (\ref{setg}). Then  the simplest solution of Eq.~(\ref{ODE}) with the repulsive
nonlinearity ($\mathcal{G}=1$)
is given by we study, the ``dark soliton" shape (\ref{ds})
 with $\eta(\br)$ substituted by $\heta(\br)$ and
valid for the
positive chemical potential ${\cal E}=\varepsilon-\nu^2>0$.

Then, according to the generalized ansatz with $\heta(\br)$, $\hvarphi(\br)$, and  $\rho(\heta)$ given by Eq.~(\ref{gcon2}), as well as $\eta(\br)$ and $\varphi(\br)$ defined in the domain ${D}_2$, we obtain the ``dark soliton"
\begin{equation}
\hat{\psi}_{ds}(\br)\!=\!\!\sqrt{{\cal E}}\rho(\heta)\tanh\!\left(\sqrt{\frac{{\cal E}}{2}}\,\heta(\br)\!\right)
\exp(i\nu e^x\!\cos y),
\end{equation}
with $\eta\equiv\eta(\br)=e^x\sin y$, which  solves  Eq.~(\ref{nls}) with  the linear and nonlinear potentials given by [c.f. Eq.~(\ref{gpotential})]
\begin{eqnarray*}
\nonumber V_{\rm ext}(\br)\!=\!\frac{e^{2x}}{8(2+\eta)^2}\!\left\{3\!-\!16(4\cE\!+\!\nu^2)
\!-\!2\eta\!\left[\nu^2\!\left(3+2\eta\right)\right.\right.
\nonumber \\
\left.\left.+2{\cE}(4\!+\!\eta)(\eta^2\!+\!4\eta\!+\!8)\right]\!\right\}, \qquad\qquad\qquad \\
g(\br) = \frac12e^{2x}(2+\eta)^3. \qquad\qquad\qquad\qquad\qquad\qquad\quad\,
\end{eqnarray*}

The presented solutions found in the domain $D_2$ behave in a way similar to the previous example, see Fig.~\ref{A_fig}. The numerical simulations prove that the found analytical solutions can indeed be used as good approximation for the stable stationary solutions of Eq.~(\ref{nls}) describing systems with physically relevant potentials.

\vspace{0.2in}

\section{Conclusion}

To conclude, we have shown that exact
analytical solutions can be obtained in a large class
of two-dimensional Gross-Pitaevskii equations with inhomogeneous
linear and nonlinear potentials, defined on
bounded domains. The method of constructing the models
is based on the properly defined conformal mapping of the given domain into a complex half-plane. In the context of applications
to Bose-Einstein condensates, the obtained solutions
having nontrivial phase depending on spatial coordinates
can be interpreted as superfluid flows. In the
case of negative scattering length (repulsive interactions)
the background flows, i.e., ones having no zeros in the
open spatial domain, appear to be stable.
The obtained results generalize previous studies devoted
to construction of the exact solutions, using the
self-similar transformation, to the two-dimensional models
given on bounded domains. Moreover, the ideas presented in this paper
are also able to apply in two-dimensional cubic-quintic models, two-dimensional multi-component models, etc.,
and to design linear and nonlinear potentials for control of
Bose-Einstein condensates and nonlinear optical fibers in limited spatial domains.

\acknowledgments

Yan was supported by the NSFC under Grant No. 11071242. VVK and AVY were supported by the grant  PEst-OE/FIS/UI0618/2011 and by   7th European Community Framework Programme under the grant PIIF-GA-2009-236099 (NOMATOS). Liu was supported by the NKBRSFC
under Grant No. 2011CB921502.



\vspace{-0.2in}

\begin{references}

\bibitem{integrable} H.-H. Chen and C.-S. Liu, Phys. Rev. Lett. {\bf 37}, 693 (1976);
 M. Bruschi, D. Levi, and O. Ragnisco, Il Nuovo Cimento A {\bf 53}, 21 (1979); R. Scharf and A. R. Bishop,
 Phys. Rev. A {\bf 43}, 6535 (1991); V. V. Konotop, O. A. Chubykalo, and L. V\'azquez,
 Phys. Rev. E {\bf 48}, 563 (1993); L. Gagnon and P. Winternitz, J. Phys. A {\bf 26}, 7061 (1993); V. V. Konotop, Theor. Math. Phys. {\bf 99}, 687 (1994).


\bibitem{Serkin}  V. N. Serkin, and A. Hasegawa, Phys. Rev. Lett. {\bf
85},  4502 (2000);  B. A. Malomed {\it et al}., J. Opt. B: Quantum Semiclass. Opt. {\bf 7}, R53 (2005); S. A. Ponomarenko and G. P. Agrawal, Phys. Rev. Lett. {\bf 97}, 013901 (2006); V. N. Serkin,
A. Hasegawa, and T. L. Belyaeva, Phys. Rev. Lett. {\bf 98}, 074102 (2007); Z. Y. Yan, Phys. Lett. A {\bf 374}, 672 (2010).


\bibitem{BEC}  Z. X. Liang, Z. D. Zhang, and W. M. Liu, Phys. Rev. Lett. {\bf 94}, 050402
(2005);  J. Belmonte-Beitia {\it et al.,}  Phys. Rev. Lett.
{\bf 98}, 064102 (2007); J. Belmonte-Beitia {\it et al.,}  Phys. Rev. Lett.  {\bf 100}, 164102 (2008); A. T. Avelar, D. Bazeia, and W. B. Cardoso, Phys. Rev. E {\bf 79}, 025602(R) (2009).


\bibitem{TGK} V. M. P\'erez-Garc\'ia, P. J. Torres, and  V. V. Konotop, Physica D {\bf 221}, 31 (2006).


\bibitem{Abdul} F. Kh. Abdullaev {\it et al}.,  Phys. Rev. E {\bf 82}, 056606 (2010).

\bibitem{YK} Z. Y. Yan and V. V. Konotop, Phys. Rev. E  {\bf 80}, 036607 (2009); Z. Y. Yan and C. Hang, Phys. Rev. A {\bf 80}, 063626 (2009); S. H. Chen and J. M. Dudley, Phys. Rev. Lett. {\bf 102}, 233903 (2009); Z. Y. Yan, V. V. Konotop, and  N. Akhmediev, Phys. Rev. E {\bf 82}, 036610 (2010); Z. Y. Yan, Phys. Scr. {\bf 78}, 035001 (2008);
    Z. Y. Yan, Phys. Lett. A {\bf 374}, 4838 (2010); Z. Y. Yan, K. W. Chow, and B. A. Malomed, Chaos, Solitons \& Fractals, {\bf 42}, 3013 (2009).

\bibitem{BYK} Yu. V. Bludov, Z. Y. Yan, and V. V. Konotop, Phys. Rev. A {\bf 81}, 063610 (2010).

\bibitem{GP} F. Dalfovo {\it et al}.,  Rev. Mod. Phys. {\bf 71}, 463 (1999).


\bibitem{Feshbach} P. O. Fedichev {\it et al}.,  Phys. Rev. Lett. {\bf 77}, 2913 (1996).

\bibitem{Amo} A. Amo,
D. Sanvitto, F. P. Laussy, D. Ballarini1, E. del Valle, M. D. Martin1, A. Lema\^itre, J. Bloch,
D. N. Krizhanovskii, M. S. Skolnick, C. Tejedor, and  L. Vi\~na
Nature, {\bf 457}, 291 (2009);
A. Amo,
T. C. H. Liew, C. Adrados, R. Houdr\'e , E. Giacobino, A. V. Kavokin  and A. Bramati
Nature Photonics, {\bf 4}, 361 (2010).

\bibitem{FT} L. D. Faddeev and L. A. Takhtajan, {\em Hamiltonian Methods in the Theory of Solitons} (Springer-Verlag, Berlin, 1987).

\bibitem{hydro} see e.g. A. J. Majda and A. L. Bertozzi, {\em Vorticity and Incompressible Flow}
(Cambridge University Press, 2002).
\end{references}
\end{document}